\documentclass[useAMS,usenatbib,preprint]{mn2e}
\usepackage{graphicx}
\def\mbf#1{\mbox{\boldmath ${#1}$}}
\newcommand{\bea}{\begin{eqnarray} }
\newcommand{\eea}{\end{eqnarray}}

\newcommand{\apj}{ApJ}
\newcommand{\mnras}{MNRAS}
\newcommand{\apss}{Ap\&Space Sci.}

\newcommand{\apjs}{ApJ Suppl.}
\newcommand{\aap}{A\&Ap}

\newcommand{\pasj}{PASJ}

\title[Instabilities of Spiral Shocks]{Instabilities of Spiral Shocks I: Onset of Wiggle Instability and its Mechanism}
\author[Keiichi Wada and Jin Koda]{Keiichi Wada,$^1$\thanks{E-mail:
wada.keiichi@nao.ac.jp} and Jin Koda,$^{1,2}$\thanks{JSPS fellow, E-mail: koda@astro.caltech.edu}  \\
$^1$ National Astronomical Observatory of Japan, Mitaka, Tokyo 181-8588, Japan \\
$^2$ Department of Astronomy, California Institute of Technology, MS 105-24, 
Pasadena, CA 91125, USA}
\begin{document}

\date{Accepted  Dec. 2, 2003, Received ; in original form }

\pagerange{\pageref{firstpage}--\pageref{lastpage}} \pubyear{2003}

\maketitle

\label{firstpage}

\begin{abstract}
We found that loosely wound spiral shocks in an isothermal gas disk caused by 
a non-axisymmetric potential are hydrodynamically unstable, if the shocks are
strong enough.  High resolution, global hydrodynamical 
simulations using three different numerical schemes, i.e. AUSM, CIP, 
and SPH, show similarly that
trailing spiral shocks with the pitch angle of larger than
$\sim 10^\circ $ wiggle, and
clumps are developed in the shock-compressed layer. 
The numerical simulations also show clear wave crests that are associated with
ripples of the spiral shocks. 
The spiral shocks tend to be more unstable in a rigidly rotating disk
than in a flat rotation.
This instability could be an origin of the secondary structures of 
spiral arms, i.e. the spurs/fins, observed in spiral galaxies.
In spite of this local instability, 
the global spiral morphology of the gas is maintained
over many rotational periods.
The Kelvin-Helmholtz (K-H) instability in
a shear layer behind the shock is a possible 
mechanism for the wiggle instability.
The Richardson criterion for the K-H stability is expressed 
as a function of the Mach number, the pitch angle, and strength 
of the background spiral potential. 
The criterion suggests that 
spiral shocks with smaller pitch angles and smaller Mach numbers 
would be more stable, and this is consistent with the numerical results.


\end{abstract}


\section{Introduction}
Spiral arms are the most prominent substructure in disc galaxies.
The spiral density wave hypothesis (Lin \& Shu 1964)
has been widely accepted
as explaining the steady stellar spiral structure. 
The spiral density waves cause shocks in the interstellar medium, 
i.e. the galactic shocks. This phenomenon  
was well studied in the 60s and 70s.
Compression of the interstellar medium 
behind the shocks is an essential mechanism in the formation of
massive stars that illuminate the spiral structure.
Assuming tightly wound spirals and steady flow, 
linear and weak nonlinear response of the gas 
in the spiral potential has been studied.
(Fujimoto 1968; Roberts 1969; Shu et al. 1972; Shu, Milione, \& Roberts 1973).

Most of these initial studies adopted rather simple assumptions for the ISM, 
i.e. isothermal equation of state and single-phase, uniform and 
non-selfgravitating gas. 
Then, more realistic cases were studied. For example, 
Sawa (1977) investigated effects of the selfgravity of the gas on the
asymptotic stational solution.
Ishibashi \& Yoshii (1984) studied an energy loss flow.
Tomisaka (1987) took into account the thermal heating 
and cooling processes for the cloud fluid.
Lubow,  Cowie, \&  Balbus (1986) studied two-component flow, i.e. isothermal gas and stars.   Among these studies, the spiral potential was assumed to be
tightly wound, and a periodic boundary condition was applied.
These studies focused on the steady state of the gas in a spiral potential,
because the observed spirals are thought to be a long-lived structure in galaxies.
Woodward (1975), on the other hand, studied gas flow in
a tightly wound spiral potential,
using time-dependent numerical simulations under the same approximation 
adopted by Roberts (1969), and showed that shocks are formed within one or two
transits of the gas through the spiral pattern.
Johns \& Nelson (1986) performed numerical
simulations on a time-dependent, 
two-dimensional gaseous response in a spiral potential, and found that
quasi-steady spiral shocks are formed.

The next question concerns the stability of the galactic shock:
Nelson \& Matsuda (1977) showed that tightly wound spiral shocks are dynamically stable. 
Dwarkadas \& Balbus (1996) discussed the linear stability of the spiral shocks assuming
a flat rotation curve and concluded that the flow is stable.
Mishurov \& Suchkov(1975) showed that the flow obtained by Roberts (1969) turns out
to be stable behind the shock front. 
On the other hand, it is unstable ahead of the shock.
Recently, Kim \& Ostriker (2002) performed two-dimensional, shearing box simulations of 
the magnetized gas in a spiral potential in an effort to understand the
origin of the secondary structure in spiral arms, namely `spurs'
(Elmegreen 1980).
They found that gaseous spurs are formed as a consequence of magneto-gravitational instabilities inside spiral arms (see also Balbus (1988) for the linear analysis).
Gravitational stability of the shocked gas in spiral arms has been well studied
in the context of the formation of the giant molecular clouds (e.g. Elmegreen 1994).

In most of the studies mentioned above, spiral shocks are assumed 
to be tightly wound, i.e. their pitch angles are smaller than 10$^\circ$, 
and a flat rotation curve which is in general suitable for the galactic outer disk is
also assumed. 
In this paper, we investigate stability of 
the gas flow in more general spiral potentials, with small ($\sim 5^\circ$) 
to large ($\sim 20^\circ$) pitch angles in various rotation curves.
We perform two-dimensional, time-dependent, global hydrodynamical simulations of 
the gas in spiral and barred potentials 
using three high-resolution numerical schemes based on different concepts.
We show that the spiral shocks generated in 
a {\it non-self-gravitating} thin disc may be dynamically {\it unstable}, 
in the sense that the shock front {\it wiggles}, and eventually knots are formed
in the shock-compressed gas (section 2 and section 3). 
The physical origin of this instability in terms of 
the Kelvin-Helmholtz (K-H) instability in a shocked layer, as well as
other implication from the numerical results will be discussed in 
section 4. A brief summary and conclusions are given in section 5. In Appendix, 
the Richardson criterion behind the spiral shock is approximately derived.

Non-linear and long-term evolution of the instability 
and comparison with observations, such as
spurs/fins in galaxies, and 
dynamics of the multi-phase, self-gravitating interstellar medium in
the spiral potential will be discussed in subsequent papers.

%
\section{Numerical Simulations}
%
\subsection{Methods}
As we will see in section 3, 
hydrodynamical simulations show that 
the spiral shocked layer is unstable in some situations.
In order to ensure that 
the instabilities are not caused by numerical artifacts,
we apply three hydrodynamical schemes based on different physical and 
numerical concepts to the same models,
i.e. an Euler mesh code (AUSM: Advection
Upstream Splitting Method), 
mesh-based semi-Lagrangian code (CIP: Cubic-polynomial Interpolation), 
and particle-based Lagrangian code (SPH: Smoothed Particle Hydrodynamics).
We find that the instability 
commonly occurs among results with these three codes.
All these schemes are standard methods in computational fluid dynamics
and astrophysical gas dynamics.
We briefly describe the three numerical schemes below:

\subsubsection{AUSM}
AUSM was developed by  Liou \& Steffen (1993), and
has been widely used for aerodynamical simulations.
AUSM is remarkably simple, but it is accurate enough based on a comparison with
 Flux-difference splitting schemes (e.g. Roe splitting) and PPM \cite{WC}.
The two-dimensional and three-dimensional versions of the scheme 
with a Poisson equation solver 
were applied to various problems of dynamics of the interstellar medium
(Wada 2001; Wada \& Norman 1999, 2001, 2002; Wada, Spaans, \& Kim 2001; 
Wada, Gerhardt, \& Norman 2002). 
We summarize the essential points of AUSM below.

The basic equations for the isothermal two-dimensional flow are written as
 \bea
\frac{\partial{\mbf{U}}}{\partial{t}}+\frac{\partial{\mbf{F}}}{\partial{x}}+\frac{\partial{\mbf{G}}}{\partial{y}}
= 0, \eea where $\mbf{U}^T \equiv (\rho, \rho u, \rho v)$,
$\mbf{F}^T \equiv [\rho u, \rho u^2+p, \rho u v]$,
$\mbf{G}^T \equiv [\rho v, \rho v u, \rho v^2 +p]$.

Flux \mbf{F} and \mbf{G} consist of two physically distinct parts,
namely advection and pressure terms:
\bea \mbf{F} = \left(
\begin{array}{l} \rho \\ \rho u\\ \rho v \end{array} \right)
u + \left( \begin{array}{l} 0 \\ p\\ 0 \end{array} \right),
\mbf{G} = \left( \begin{array}{l} \rho \\ \rho u\\ \rho v
\end{array} \right) v + \left( \begin{array}{l} 0 \\ 0\\ p
\end{array} \right). \eea
In AUSM, these two terms are separately
split at a cell surface. 
The van Leer-type flux splitting process is applied separately to
these two terms (Liou 1996). 

We achieve third-order spatial accuracy with the Monotone
Upstream-Centered Schemes for Conservation Laws (MUSCL) approach  
\citep{VL}.  To satisfy the TVD (Total Variation Diminishing)
 condition using MUSCL, we introduce a limiting function.
The code was checked for various test problems, such as
one-dimensional and two-dimensional shock tube problems,
collision of strong shocks, reflection of shocks at
a wall, and propagation of 2-D blast waves. See details in Wada \& Norman (2001).
We use a Cartesian grid with $128^2-2048^2$ zones.

\subsubsection{CIP}

The Cubic-polynomial Interpolation or Cubic Interpolated Propagation (CIP) 
method, developed by Yabe et al. \citep{yabe91}, 
is a kind of semi-Lagrangian method, in which 
the hyperbolic equation $\partial \mbf{f}/\partial t + (\mbf{u} \cdot \Delta) \mbf{f} 
= \mbf{g}$, where e.g. $\mbf{f} \equiv (\rho, \mbf{u})$ and
$\mbf{g} \equiv (-\rho \Delta \cdot \mbf{u},
-\Delta p/\rho$), is split into advection and non-advection phases. 
The former is advanced by using a profile interpolated with
a cubic polynomial, and the profile is determined from the
time evolution of the quantity $\mbf{f}$ and the spatial derivative $\Delta \mbf{f}$.
The CIP method is a stable and less diffusive scheme for compressible and incompressible
flows with sharp boundaries, and it has been 
applied to various problems in the field of 
computational fluid dynamics and astrophysics (e.g., Kudoh, Matsumoto \& Shibata 1998 for formation of astrophysical jets).
One remarkable feature of this code is that discontinuity is well resolved with
relatively small numbers of grid points.
The code used here has first-order accuracy in space and time.
The number of grid zones used here is the same as that which runs with AUSM.

\subsubsection{SPH}

We use the Smoothed Particle Hydrodynamics (SPH) code
presented by Koda \& Wada (2002), which basically
uses Benz's formulation (Benz 1990), 
using the spline kernel by Monaghan \& Lattanzio (1985) with the modification for its gradient (Thomas \& Couchman 1992).
The correction term for viscosity is taken into account to 
avoid large entropy generation in pure shear flows (Balsara 1995).
The SPH smoothing 
length $h$ varies in space and time, keeping the number of particles within
the radius of $2h$ at an almost constant of 32
according to the method of Hernquist \& Katz (1989).
The leapfrog integrator is adopted to update positions and velocities.

In order to represent an initial uniform-density gas disc, we randomly
distribute $10^5$ gas particles in a 2-D disc with a radius of 3 kpc.
The gas particles follow pure circular rotation that balances the
centrifugal force.
\subsection{Basic equations and Boundary conditions}

We solved the following equations numerically in two-dimensions:
\begin{eqnarray}
\frac{\partial \Sigma_g}{\partial t} + \nabla \cdot (\Sigma_g \mbf{v}) &=& 0, \label{eqn: rho} \label{eq1} \\ 
\frac{\partial \mbf{v}}{\partial t} + (\mbf{v} \cdot \nabla)\mbf{v}
+\frac{\nabla p}{\Sigma_g} &=& -\nabla \Phi_{\rm ext}  , \label{eqn: rhov}  \label{eq2} 
\end{eqnarray}
where, $\Sigma_g,p,\mbf{v}$ are surface density, pressure, and velocity of the gas.
$\Phi_{\rm ext}$ is a fix external potential.
We assume the isothermal equation of state, and gas temperature is $10^4$ K, 
unless otherwise mentioned.

In AUSM and CIP simulations, the boundary conditions are set as follows.
In ghost zones at boundaries, the physical
values remain at their initial values during the calculations. From
test runs, we found that these boundary conditions are much better than
other boundary conditions, such as `outflow' boundaries, which cause strong
unphysical reflection of waves at the boundaries.
Even though this boundary condition does not seem to affect 
the gas dynamics in the computational box, 
we put the boundaries far enough
from the region where we study the stability of the spiral shocks
to avoid unexpected numerical artifacts. The size of the computational
box is typically 8 kpc $\times$ 8 kpc (model A, B, and D, see below) 
or 32 kpc $\times$ 32 kpc (model C).  In SPH, we use a free boundary.
The simulations are performed in a rotating frame of a fixed pattern 
speed of the non-axisymmetric potential.
We assume initially uniform surface density ($7 M_\odot$ pc$^{-2}$),

\subsection{Potential models}
\begin{figure}
   \begin{center}
 	\includegraphics[width = 7cm]{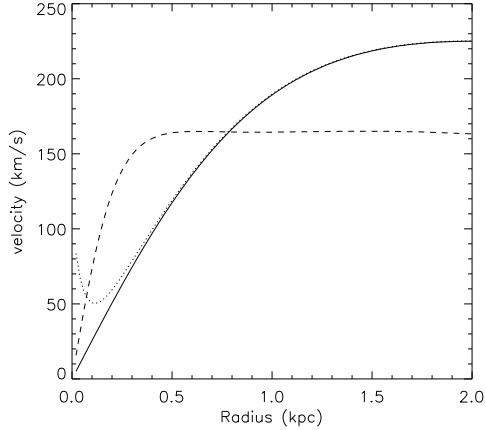}
        \caption{Rotation curves adopted in the numerical simulations.
Solid line (model A): nearly rigid rotation models, dashed line: nearly flat rotation curve (model B), and dotted line (model D): the bar model with a central massive black hole. }
\label{fig: rot_curv}

   \end{center}
  \end{figure}

\begin{figure}
   \begin{center}
        \includegraphics[width = 7cm]{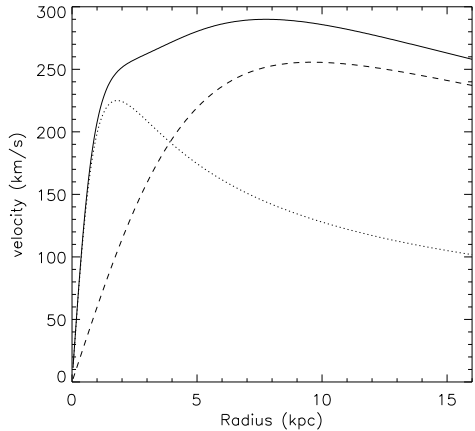}
        \caption{The rotation curve of model C. Solid line: $\Phi_0+\Phi_2$,
dotted line: $\Phi_0$, and dashed line $\Phi_2$.}
\label{fig: model-c}

   \end{center}
  \end{figure}

We solve two-dimensional gas dynamics in a 
time-independent, non-axisymmetric external potential.
Four different potential models are used; 
three spiral potentials (model A, B, and C) and one bar potential (model D).
One of the spiral models (model C) is assumed a large scale disk 
(initial radius is 16 kpc), and others and the bar model are for 
inner galactic disks  (initial radii are 4 kpc).
Rotation curves derived from the axisymmetric potentials 
of model A, B, and D, are shown in Fig. \ref{fig: rot_curv}, and
model C in Fig. \ref{fig: model-c}.
The four models are defined as below:
\bea
{\rm model\;\;A} &:& \Phi_{\rm ext} \equiv \Phi_0(R) + \Phi_1(R,\phi), \\
{\rm model\;\;B} &:& \Phi_{\rm ext} \equiv \Phi_0(R) + \Phi_2(R) + \Phi_3(R,\phi), \\
{\rm model\;\;C} &:& \Phi_{\rm ext} \equiv \Phi_0(R) + \Phi_2(R) + \Phi_3(R,\phi), \\
{\rm model\;\;D} &:& \Phi_{\rm ext} \equiv \Phi_0(R) + \Phi_4(R) +  \Phi_1(R,\phi),
\eea
where the axisymmetric components $\Phi_0, \Phi_2,$ and $\Phi_4$ 
are defined as
\bea
 \Phi_0(R) &\equiv& {a v_a^2 (27/4)^{1/2}}{(R^2 + a^2)^{-1/2} }, \\
 \Phi_2(R) &\equiv& {b v_b^2 (27/4)^{1/2}}{(R^2 + b^2)^{-1/2} }, \\
 \Phi_4(R) &\equiv& {c v_c^2 (27/4)^{1/2}}{(R^2 + c^2)^{-1/2} }, 
\eea
and the constants $a, b,c,v_a, v_b$ and $v_c$ in the four models 
are summarized in Table 1.
The non-axisymmetric components (i.e. spiral and bar potentials),
$\Phi_1$ and $\Phi_3$ are given by 
\bea
 \Phi_1(R,\phi) &\equiv& \varepsilon_0 \frac{a R^2 \Phi_0}{(R^2 + a^2)^{3/2} } 
\cos[2\phi + 2\cot i \cdot \ln (R/R_0)], \\
 \Phi_3(R,\phi) &\equiv& \varepsilon_0 \frac{b R^2 \Phi_2}{(R^2 + b^2)^{3/2} } 
\cos[2\phi + 2\cot i \cdot \ln (R/R_0)]
\label{eq: non-axi}
\eea
where $i$ is the pitch angle of 
the spiral potential (see Fig. \ref{fig: fig4}),
and $R_0$ is an arbitary constant ($R_0 = 0.9$ kpc is
used for all models). A constant $\varepsilon_0$ represents maximum strength of 
the non-axisymmetric potential.
For the bar potential model D, $i=\pi/2$ is assumed with $\Phi_1$.

\begin{table}
\begin{center}
\caption{Parameters for the four potential models. Units of the core radii are
kpc, and km s$^{-1}$ for the velocity $v_a$, $v_b$, and $v_c$.}
\begin{tabular}{c|cccccc}\hline
model  & a & b & c & $v_a$ & $v_b$ & $v_c$ \\ \hline
A & 1.4 & -- & -- & 220 & -- & -- \\
B & 0.2 & 1.7 & -- & 150 & 103 & -- \\
C & 1.3 & 6.8 & -- & 220 & 250 & -- \\
D & 1.4 & -- & 0.01 & 220 & -- & 311 \\ \hline
\end{tabular}
\label{tab: tab0}
\end{center}
\end{table}

Model D is chosen in order to see stability of spiral shocks formed by 
a different mechanism, i.e. resonance-driven spiral shocks.
We add a point mass at the galactic centre ($\Phi_4$),
which resembles, for example,  a supermassive black hole.
In model D, the `black hole' mass is assumed to be 1\% of the total dynamical mass.
This causes a `nuclear' inner Lindblad resonance (nILR), and 
it is known that trailing spiral shocks are formed around the nILR 
(Fukuda, Wada, \& Habe 1998). 
This characteristic resonant-driven structure also appears around the outer ILR (Athanassoula 1992; Piner et al. 1995). 
Using two-dimensional SPH simulations,
Fukuda, Habe, \& Wada (2000) revealed that 
the trailing spirals can be gravitationally unstable, and claimed that
this process is important in fuelling the gas to the galactic centre.
Therefore, stability of such trailing spirals in the galactic nuclear region
in cases in which the gas is not self-gravitating is worth studying.
We can also study the generality of the stability/instability of the
spiral shocks.

%
\section{Numerical Results}
%

\label{section: result}
%
%
\label{section: spiral_pot}
  \begin{figure}
   \begin{center}
 	\includegraphics[width = 9cm]{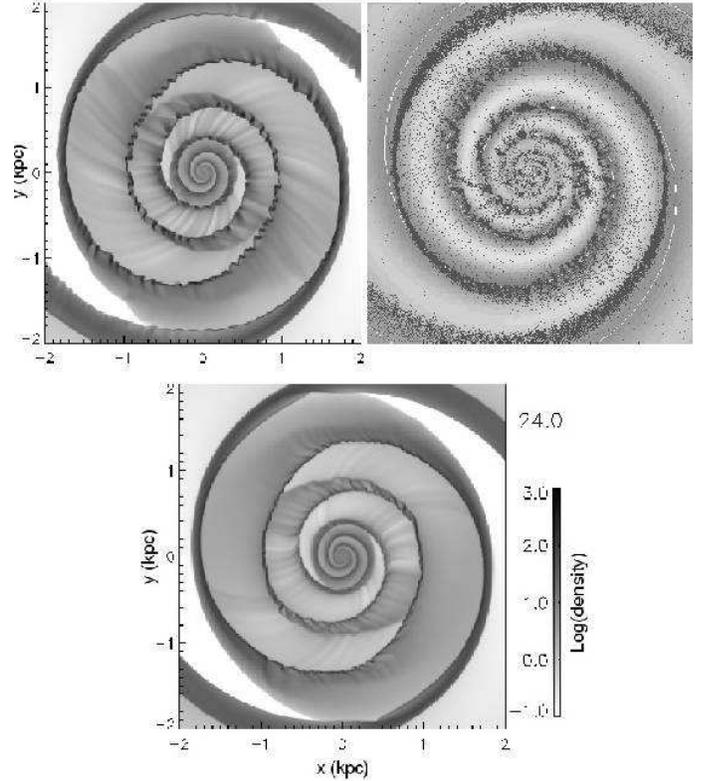}
        \caption{Response of the gas disc to a spiral potential (model A with 
$i =10\degr$, $\varepsilon_0 = 0.1$, $\Omega_p = 26$ km s$^{-1}$ kpc$^{-1}$) 
at $t =24$ Myr. 
(a) Density distribution given by AUSM with $1024^2$ cells. Gray-scale represents 
log-scaled surface density with a unit of $M_\odot$ pc$^{-2}$. (b) SPH with $10^5$ particles (roughly there are $4\times 10^4$
 particles in this box).
 Gray-scale represents amplitude of the spiral potential, and the white line is 
a bottom of the potential. (c) Same as (a), but by CIP. }
\label{fig: result_1}

   \end{center}
  \end{figure}

In Fig. \ref{fig: result_1}, we show typical numerical results obtained by
the three different numerical schemes, i.e. SPH, AUSM, and CIP, explained
in section 2.1.
`Wiggle' instabilities of the spiral shock, namely ripples of the shock fronts
and less dense fin-like structure are clearly seen in all the results.
The global morphology of the spirals in all results
is quite similar. Comparing Fig. \ref{fig: result_1}(a) with \ref{fig: result_1}(c),
one may find that the instability is less prominent in the result by CIP.
This is because the CIP code used here is less accurate (i.e. first-order scheme) than AUSM (spatial third-order scheme), 
and therefore the shock is captured less sharply.
Although the SPH result is noisier and seems to
already be in a non-linear phase especially in the inner region,
comparing to the other two results obtained by the mesh-based codes,
wavelengths of the instability are
similar between the models.
The fin-like structures seen in the inter-arm regions are density waves
originating in the wiggle instability of the shock front. This
phenomenon will be
discussed in detail in a subsequent paper (see also section 5).

  \begin{figure}
   \begin{center}
        \includegraphics[width = 9cm]{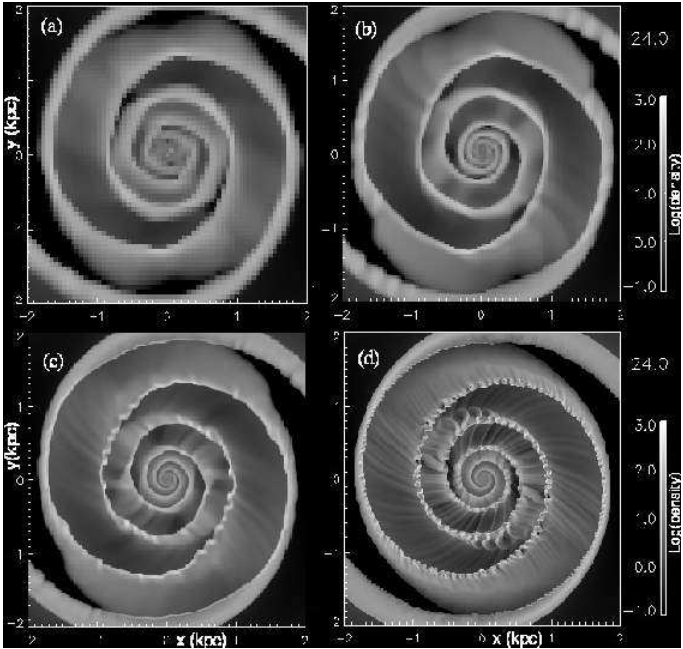}
        \caption{Effect of changing a spatial resolution on the instabilities.
The potential model is the same as that in Fig. \ref{fig: result_1}.
(a) Density distribution of a central 1/4 region of 
the computational box with $128^2$ cells at $t=24$ Myr. (b), (c) and (d):
Same as (a), but with $256^2$, $512^2$, and $2048^2$, respectively.
The parameters used here are the same as those for models in Fig. 3.}
\label{fig: result_3}

   \end{center}
  \end{figure}

For further discussion, we use results with the AUSM scheme. 
The four panels of Fig. \ref{fig: result_3} show
how the numerical results are affected by
the spatial resolution. Although global morphology of the spirals is similar 
in all results, the wiggle instability of the shock fronts 
is not clear in the results with $128^2$ and $256^2$ cells. 
We need at least $512^2$ grid points to resolve the instability in this case.
Fig. \ref{fig: result_3} (d) shows that characteristic scale of the
ripples is roughly the width of the shock-compressed layer, which is
about 100-200 pc. We can also see that spur-like structures are
developed into the inter-arm regions, which are 
associated with the ripples (see also Fig. \ref{fig: result_31}).


  \begin{figure}
   \begin{center}
        \includegraphics[width = 5cm]{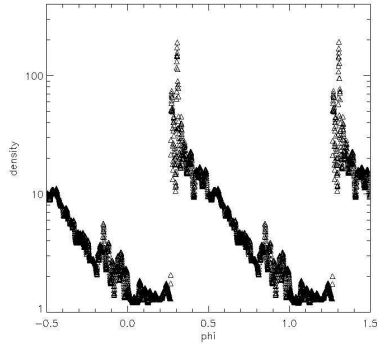}
        \caption{Azimuthal density profile at 
$R=1.5$ kpc of Fig. \ref{fig: result_3} (d). The unif of 
density is $M_\odot$ pc$^{-2}$.}
\label{fig: result_phiden}
   \end{center}
  \end{figure}

From the azimuthal density profile (Fig. \ref{fig: result_phiden}), 
it is clear that reveals sharp shocks where density jump is $\sim $ 100, and
large density fluctuation in the shock compressed layer,
which is originated from the clumps seen in Fig. \ref{fig: result_3}, are 
formed.  One may also notice density fluctuation
in the inter-arm regions. This can be seen as many `crests' in
Fig. \ref{fig: result_3} (d).

  \begin{figure}
   \begin{center}
        \includegraphics[width = 9cm]{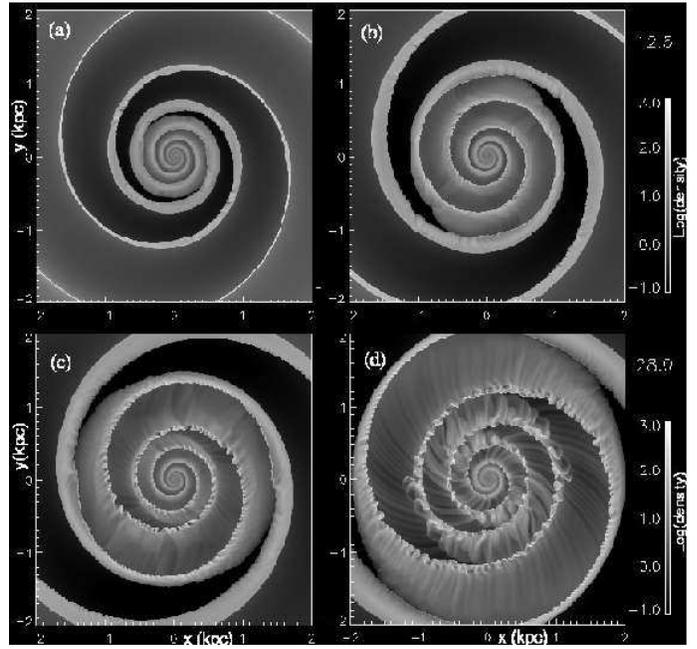}
        \caption{Evolution of the instability in the same 
model shown in Fig.  \ref{fig: result_3} (d). 
Four snapshots are at (a) t=7.5 Myr, 
(b) 12.5 Myr, (c) 17.5 Myr, and (d) 28 Myr.}

\label{fig: result_31}

   \end{center}
  \end{figure}

Figure \ref{fig: result_31} shows time evolution of the spiral shocks in 
the same model of Fig. \ref{fig: result_3} (d).
The wiggle instability begins in the inner region first, then
the outer shocks becomes unstable. At $t=28$ Myr (d),
the instability is in a non-linear stage, and `spurs' are developed
in the inter-arm regions towards the other spiral shocks.
Typical time-scale of the linear growth of the instability is $\sim 10$ Myr
at $R \sim 500$ pc and $\sim 30 $ Myr at $R \sim 2$ kpc. 
This is about a half of the orbital time-scale.
The spurs are nearly regulary spaced, and their intervals seem to be
larger in the non-liear phase than in the early phase.

  \begin{figure}
   \begin{center}
 	\includegraphics[width = 9cm]{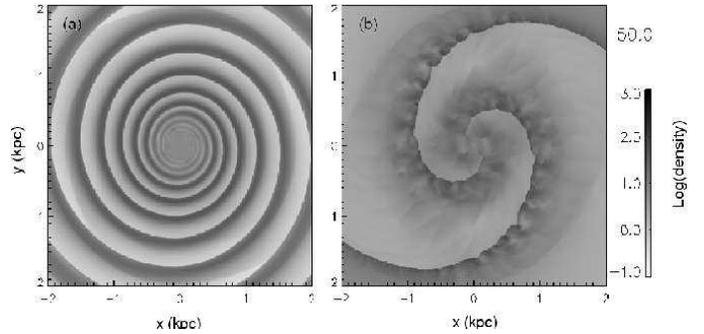}
        \caption{Response of the gas disc to a spiral potential (model A with 
$i =5\degr$ and $i=20\degr$).  $\varepsilon_0 = 0.1$, $\Omega_p = 26$ km s$^{-1}$ kpc$^{-1}$ and $T_g  = 10^5 $ K at 50 Myr. $1024^2$ cells are used in both models.}
\label{fig: result_2}

   \end{center}
  \end{figure}

The two panels in Fig. \ref{fig: result_2} demonstrate
 that the stability of the 
shock fronts is 
sensitive to the pitch angle of the spiral potential (model A). It shows that 
the shocks are stable for $i = 5\degr$, whereas they are unstable for 
more loosely wound spirals, i.e. $i=20\degr$. 

\begin{figure}
   \begin{center}
 	\includegraphics[width = 5cm]{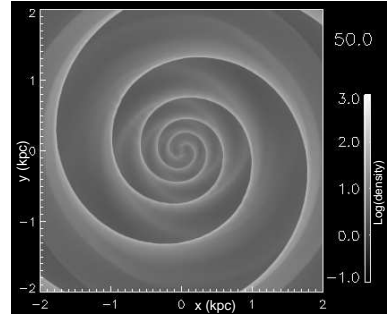}
        \caption{Same as Fig. \ref{fig: result_2}, but for the case that 
the rotation curve is nearly flat (model B) with $\varepsilon_0 = 0.03$. }
\label{fig: result_4}

   \end{center}
  \end{figure}

We found that if the rotation curve is flat (model B), the spiral shocks are  
stable (Fig. \ref{fig: result_4}), provided that 
the spiral potential is weak (but strong enough to
produce the shocks).
Note that even if the unperturbed rotation curve is flat,
the perturbed potential can cause a radial gradient in the circular velocity
on a local scale. This local velocity gradient 
is essential for the onset of the instability
as discussed in section 4 and Appendix.

%
%

\begin{figure}
   \begin{center}
        \includegraphics[width = 7cm]{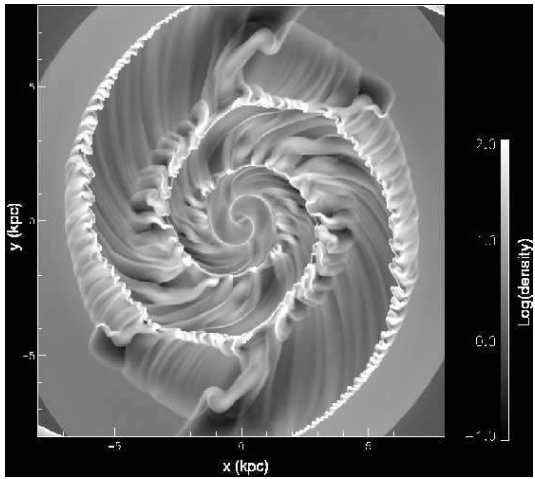}
        \caption{Density distribution in a spiral potential (model C) with
$i = 15^\circ$ at $t=10^8$ yr. $2048^2$ cells are used.}
\label{fig: big_spiral} 

   \end{center}
  \end{figure}

We found that the wiggle instability of the shocked layer appears not only 
in the central kpc of galaxies, but also in a larger size disk.
Fig. \ref{fig: big_spiral} is a snapshot of model C, in which 
a disk with four times larger size than that in model A and B is used.
The instability and spur-structure in the inter-arm regions are
similar between the model A and C, suggesting that the wiggle 
instability can occur on any scales.

%
%

  \begin{figure}
   \begin{center}
        \includegraphics[width = 8cm]{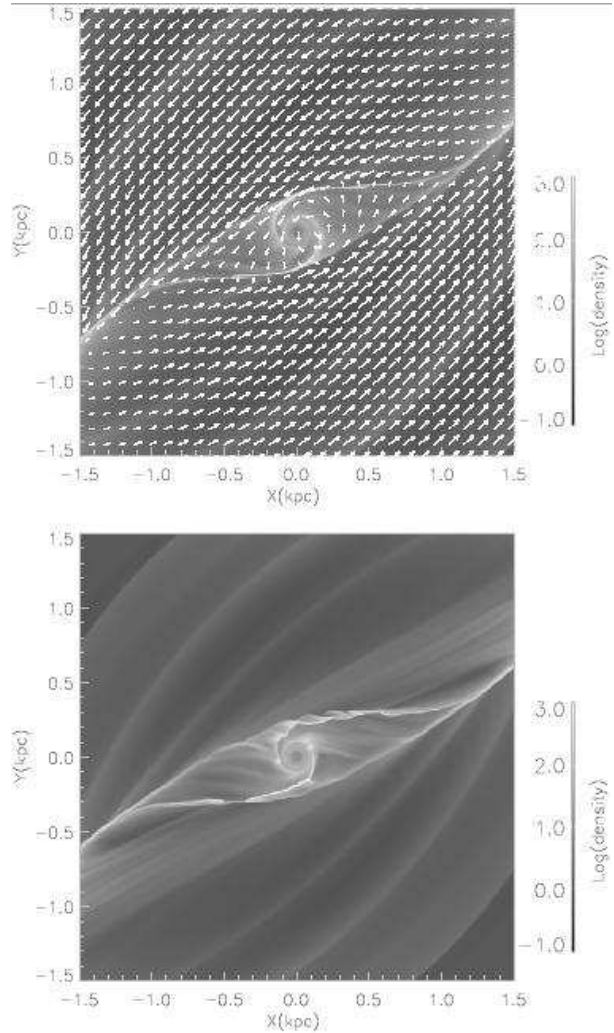}
        \caption{Spiral shocks generated around the nuclear Lindblad resonance.(
model D with $\varepsilon_0 = 0.1$, $\Omega_p = 30$ km s$^{-1}$ kpc$^{-1}$) at $t=$ 60 (upper panel) and 70 Myr (lower panel). Gray-scale is the log-scaled density, and vectors represent a velocity field.}
\label{fig: bar_result}

   \end{center}
  \end{figure}

Finally, we examine whether the instability seen in the previous 
section is a phenomenon peculiar to
the spiral potential. Fig. \ref{fig: bar_result}
is the density and velocity field near the nuclear Lindblad resonance caused by
a central mass concentration, such as a supermassive black hole, in 
a weak bar potential (model D). In this case, off-set straight shocks are 
formed outside the nuclear spiral shocks. 
We found that 
the shocks are unstable, and they wiggle as seen in the snapshot at $t=70$ Myr.
One may also notice the wavelets in the post-shock regions.

%
\section{Discussion}
%

  \begin{figure}
   \begin{center}
 	\includegraphics[width = 8cm]{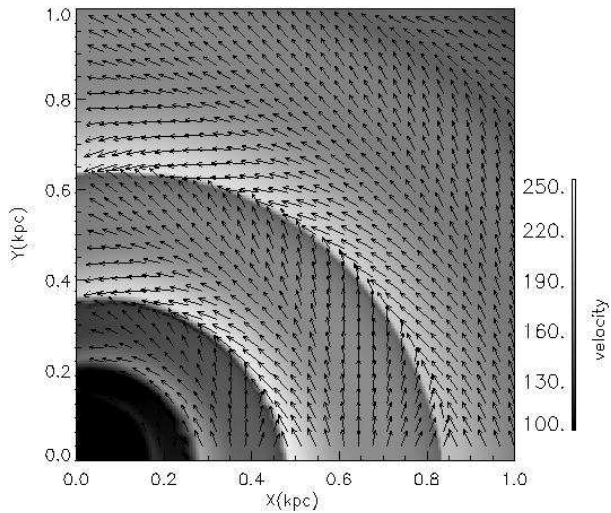}
        \caption{Velocity distribution in 
the early phase of the evolution for model A with
$i=10\degr, \varepsilon_0 = 0.1$.
 Gray-scale represents amplitude of the velocity.}
\label{fig: result_vel}
   \end{center}
  \end{figure}

The results shown in section 3 strongly suggest 
that the `wiggle instability'
 of the shocked layers followed by formation of clumps and spurs 
is not caused by numerical artifacts, 
but it has a physical origin.
We suspect that 
they originate in the Kelvin-Helmholtz (K-H) instability.
As the velocity fields  in Fig. \ref{fig: bar_result} and \ref{fig: result_vel} show, there is a strong velocity shear behind the shock. 
In standing spiral shocks caused in a rotating disk, 
like galactic shocks or bar-driven shocks, it is most likely that 
velocity gradient is generated behind the shocks.
In general,
the K-H instability can develop in a contact surface
of two fluid layers with different tangential velocities, even if
the velocity difference is small, provided that there is no surface tension.
However, in the situation considered here, buoyancy force can suppress
the instability, because there is also a density gradient behind the shock
under the spiral (or bar) potentials.
The Coriolis force could also affect development of the K-H instability 
(Chandrasekhar 1961). However, for the sake of simplicity,
we ignore this kind of effect in the following discussion.
%
%


  \begin{figure}
   \begin{center}
 	\includegraphics[width = 8cm]{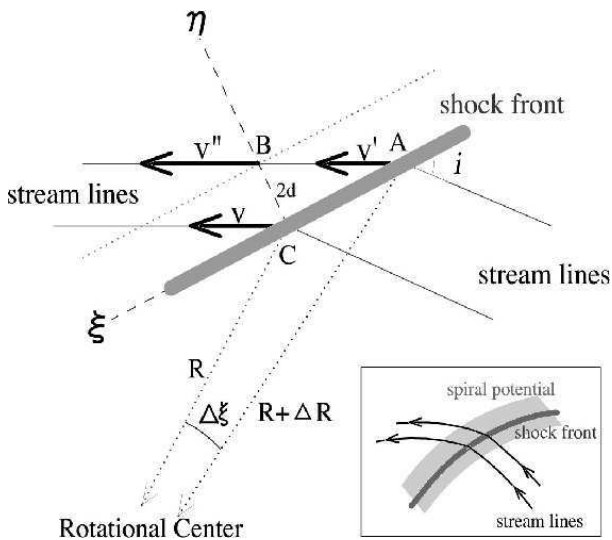}
        \caption{Spiral shock and streamlines, for
which we assumed that the curvature can be negligible, and 
therefore the spiral shocks are modeled as oblique shocks. 
For $\Delta R \ll R$, the 
spiral shock can be approximated as an oblique shock. A streamline
is changed its direction at point A with velocity $v'$, and it is 
accelated toward the other shock. At point B, its velocity is $v''$. 
A velocity gradient is expected between points B and C. A characteristic
scale of the shocked layer is $2d$. ($\xi, \eta$) is a orthogonal coordinate
along the spiral with the pitch angle, $i$. }
 \label{fig: fig4}
   \end{center}
  \end{figure}

A condition of the K-H {\it stability} in a sheared layer with a density gradient 
is characterized 
using a non-dimensional number, i.e. the Richardson number, 
$J$, defined by
\bea
J \equiv - \frac{g}{\rho} \frac{d\rho/dz}{(d u/dz)^2},
\label{eq: richard_00}
\eea
where $g$ is gravitational acceleration, which is normal to the shock front ($z$-direction), and $u$ is shear velocity.
$J$ measures the ratio of the buoyancy force to the inertia, and 
a {\it necessary condition} for the K-H {\it stable} 
is that  $J > 1/4$ (e.g., Chandrasekhar 1961). 
Here we give a rough estimate of $J$ behind the shock.
Figure \ref{fig: fig4} shows flows passing a spiral shock
schematically. The K-H instability could be originated in
the velocity shear ($\Delta u$) between position B and C, i.e. $v''$ ($> v'$) and $v$.
Equation (\ref{eq: richard_00}) is  then $J \approx 2 g d/\Delta u^2$ for the shocked layer with
a width of $2d$.
Then using
\bea
g \approx \frac{\partial \Phi_{\rm na}}{R\partial \eta} \approx \frac{\varepsilon v_\phi^2}{R \sin i},
\eea
where $\Phi_{\rm na}$ is a non-axisymmetric component of the potential (e.g. $\Phi_1$ or
$\Phi_3$ in section 2.3), $\varepsilon$ represents a fraction of
the non-axisymmetric part to the axisymmetric potential, and $v_\phi$ is circular velocity, 
and 
\bea
\Delta u \approx (v''-v)\cos i \approx v_\phi \left(1- \frac{1}{M^2}\right) \cos i,
\eea
where $M$ is the pre-shock Mach number, 
the Richardson number can be expressed as a function of 
the Mach number and the pitch angle: 
\bea
J \approx \frac{2\varepsilon d}{R \sin i (1- \sin^2 i) (1-1/M^2)^2}.
\label{eq: richard_5}
\eea
This suggests that for a given $M$, 
a larger $J$ is obtained for smaller $i$, and 
for a larger $M$, $J$ is smaller.
For example, for $\varepsilon = 0.1$ , $M = 10$, and
$d/R = 0.1$, $J \sim 0.12 $ and 0.39 for 
$i=10^\circ$ and $i= 3^\circ$, respectively.
Although one cannot predict whether the flow is unstable
only from the Richardson criterion, 
eq. (\ref{eq: richard_5}) 
suggests that tightly wound spiral shocks may be relatively stable, 
comparing to open spirals.
This tendency is in fact confirmed by our numerical simulations 
(e.g. Fig. \ref{fig: result_2}). 
More detail treatment of the flow and derivation of 
the Richardson number behind shocks is given in 
Appendix, where Fig. \ref{fig: richard} shows how 
$J$ depends on the Mach number and the pitch angle, 
which is qualitatively the same sense of eq. (\ref{eq: richard_5}).


As the ripples develop in the shock compressed layers,  many spur-like structures, or ``crests'', which are associated with the clumps, are formed behind the shock,
 and they extend to the inter-arm regions with
a large pitch angle.
These structures are morphologically similar to the dust-lanes observed in spiral galaxies. 
A typical example of this is the central regions of M51 (Scoville et al. 2001, see also Elmegreen 1980). 
Kim \& Ostriker (2002) argued that those spur-like structures are formed 
as a consequence of magneto-Jeans instability using two-dimensional, local
MHD simulations. 
The wiggle instability we found in the numerical simulations would be
an another mechanism to form 
the spurs, which originates in the K-H instability of the shock-compressed layer.
The mechanism of forming spurs is following.
The clumps formed by the instability have an internal velocity/angular momentum
gradient that causes a phase shift in the orbital motion, 
and results in deformation of the clumps. Hence, a part of each clump
with relatively small angular momenta falls along the spiral shocks, 
whereas the gases with larger angular momenta try to 
maintain nearly circular rotation.
As a result they move away from the shock toward the other spiral shock.
Because of 
the internal gradient of the angular momentum of a clump formed behind a shock,
the clump gas eventually extends 
to the inter-arm region as it rotates in
the galactic potential.   One should note that even in a highly nonlinear phase,
the global galactic ``shocks'' are still present, 
although they are no longer continuous shocks. 
In subsequent papers, we will discuss the non-linear development of
the ripple instability of spiral shocks and the formation mechanism of the spurs.
Detail comparison with observations would be also necessary.

In the present simulations, we ignored self-gravity of the ISM.
Self-gravity is essential for dynamics and structure of the ISM in galaxies,
especially for forming high-density regions and star formation. This is 
more prominent under the non-axisymmetric external potential (Wada \& Koda
2001). As shown in Fig. \ref{fig: result_phiden}, spiral shocks produce
compressed layers where the density is a factor 100 larger than that 
in the pre-shock region. This is because we assumed isothermal equation of
state for the ISM, and the Mach number is $\sim 10$. Although this density 
contrast is comparable to a ratio
between a typical density of giant molecular clouds and the average ISM density
in a galactic disk, inclusion of self-gravity with realistic cooling processes
is necessary to discuss hydrodynamical instabilities in real galaxies.
Moreover, it is of interest to study how 
the gravitational instability of the shocked layer (Balbus \& Cowie 1985 and Balbus 1988)
relates with the wiggle instability.
Even if the post-shock layer is gravitationally stable, 
self-gravity would trigger collapse of clumps formed by the wiggle instability, and
it could lead star formation.

Another interesting phenomenon related with the wiggle instability is
interstellar turbulence. In a non-linear phase, clumps and spurs interact each other,
and also collide with other spiral shocks. As a result, irregular motion of the
gas is generated, and it could develop turbulent motion of the ISM.

One might think that our findings seem inconsistent with previous 
studies on spiral shocks. For example, Dwarkadas \& Balbus (1996) 
claimed that the spiral shocks are K-H stable, using a linearized perturbation
analysis of the post-shock flow. In their analysis, they numerically integrate
a set of linearized perturbed equations assuming a tightly wound spiral
(Roberts 1969). They found that the amplitude of the perturbation increases
initially for about one-third of a rotation period, and then it decreases by a
factor of 20-50 from its maximum value toward a steady state within 
one rotation period. It should be noted, however, that they assumed 
a tightly wound spiral in a flat rotation
curve ($\kappa = 2 \Omega$), which is a preferable condition for
the K-H stable as shown in 
the numerical results and analysis in Appendix.
The other point is that 
one-third of a rotation period would be long enough to form the ripples,
clumps and spurs.
These sub-structures found in our numerical
simulations are not steady, but rather, temporal from
a Lagrangian point of view. However, successive formation
and destruction of these substructures could keep global, long-lived 
non-axisymmetric, non-uniform features in the interstellar medium 
in spiral galaxies.

The instability that we found is not apt to appear in
tightly wound spirals with a flat rotation curve. Therefore 
it is not surprising that the instability has not been found in the many studies in 
the 60s and 70s on spiral shocks. 
Yet there have been some pioneering studies on 
the time-dependent, two-dimensional flow in a spiral potential with various
pitch angles.
Johns \& Nelson (1986) performed such simulations using two distinct
codes, i.e. a Eulerian grid code and SPH. Unfortunately their spatial resolution
was rather limited ($63^2 \times 2$ polar grid), and as we showed in 
section 3, this is not fine enough to resolve the instability.
However, interestingly, they noticed
secondary structures in their spiral arms. The spiral arms in their
simulations are not smooth, 
and local density peaks are seen in the contour maps.
They mentioned a possibility of a physical origin of 
the modulation of the main two-armed response rather than
some stochastic or numerical effect, however they did not confirm it.
The secondary structure that they found might come from the
instability of spiral shocks discussed in this paper.

\section{Conclusions}

Using high-resolution numerical simulations, 
we have found that spiral shocks caused by a galactic spiral potential and
by bar-driven resonances 
can be unstable, in the sense that shocked layers are rippled, and in a non-linear
phase they fragment into clumps followed by forming spurs.
The instability is most likely caused by the Kelvin-Helmholtz instability in the 
post-shock layers where density gradient and velocity shear are present under the influence of 
the gravity of the spiral potential.
This instability appears in spiral shocks on 
small ($R <$ kpc) to large ($\sim 10$ kpc) scales, also suggesting that local velocity shear behind the shocks drive it. 
Spiral shocks tend to be more unstable, 
if the their pitch angle is larger ($ \ga 10^\circ$), and for a higher Mach number ($M \ga 3$).  
The numerical results are not inconsistent with the discussion using the Richardson criterion (section 4 and Appendix). However, one should be careful that the Richardson 
criterion itself is just a necessary condition for the K-H stability, and even if
$J < 1/4$ the flow could be stable. In this sense, 
the K-H instability is one possible mechanism of the wiggle instability.
We would need careful linear analyses for perturbations in loosely wound spiral shocks
for further discussion.

\section*{Acknowledgments}
We thank H. Fukuda for his contribution on the CIP code.
We are also grateful for N. Scoville, R. Wyse, W. Kim, and Y. Sofue for their
comments and suggestions. The anonymous referee also gave us many important 
suggestions. K. W. was supported by a Grant-in-Aid for Scientific Research (no. 15684003).
J. K. was supported by the Japan Society for the Promotion of 
Science (JSPS) for Young Scientists.
A part of numerical works were performed on facilities in 
Astronomical Data Analysis Center, National Astronomical Observatory of 
Japan.

\section*{Appendix: Richardson criterion behind a spiral shock}

In this Appendix, 
we stand on a hypothesis, i.e. the wiggle instability found in
the numerical experiments is caused by the  Kelvin-Helmholtz
instability. 
We give the Richardson number behind the
spiral shocks as a function of 
parameters, such as 
the Mach number and the pitch angle of the spiral potential.
This is useful to evaluate whether the sheared 
flow can be Kelvin-Helmholtz stable.
One should note, however that 
we simplify the kinematics of the gas in spiral potential, and 
the Richardson criterion here is not derived from a dispersion relation
in a sheared layer caused by open spiral shocks,
which may be ultimately given
in a future work (see Dwarkadas \& Balbus (1996) for linear analysis of
tightly wound spiral shocks in a flat rotation curve).
Yet a simple `toy model' below would be useful to understand physics 
behind the numerical results. 

We simply assume almost straight streamlines behind the shock 
and $\Delta R/R \ll 1$  (Fig. \ref{fig: fig4}) as we did in section 4.
We also assume that the flow is isothermal, and
the asymptotic solutions of the flow in a spiral potential is
used to have the velocity and density gradient in the post-shock layer.
First we consider only velocity shear which is parallel to
the shock as a source of the K-H instability. 
Effect of the perpendicular velocity, $v_\xi$, will be considered
later.

Under these assumptions, the Richardson number [eq. (\ref{eq: richard_00})]
 can be expressed as 
\bea
J &\approx& g \sigma d/u_0^2, 
\label{eq: richard_0}
\eea
with the velocity and density differences in a shocked layer,
i.e.  $u_0^2 \equiv (v_{\rm C} - v_{\rm B})^2/4 $ and 
 $\sigma \equiv ({\rho_{\rm C} - \rho_{\rm B}})/({\rho_{\rm C} + \rho_{\rm B}})$ (B and C denote positions shown in Fig. \ref{fig: fig4}), and
$2d$ is width of the shock layer.

Width of the shock layer, $2d$, is related to the Mach number $M$ 
(i.e. $v_\phi/c_s \sin i$, where
$v_\phi$ is the pre-shock velocity on a rotating frame with a pattern speed $\Omega_p$) 
and to the pitch angle $i$ as
\bea
2d = \frac{\Delta R}{\sin i} \tan i/M^2 
= \frac{\Delta R}{M^2 \cos i}.
\label{eq: d}
\eea
As shown in Fig. \ref{fig: fig4}, a flow passing the oblique shock has
velocity $v'$ at position A, and is accelerated toward the other spiral shock.
As a result, a velocity gradient is generated behind the shock, for example
between positions B and C (i.e. $v_{\xi 0}$ and $v_{\xi 0}''$. Subscript 
``0'' denotes unperturbed variables).
This velocity gradient, which is normal to the spiral shock, may be a source of the K-H instability. 
The post-shock velocity transverse to the shock, $v_{\eta 0}$,
is $v_{\eta 0} (R) \equiv  v_\phi (R) \sin i /M^2 $,
and the post-shock velocity parallel to the shock is
 $v_{\xi 0} (R) \equiv v_\phi (R) \cos i$, and 
the post-shock velocity at position A is  $v' (R) = \sqrt{v_{\xi 0}^2 + v_{\eta 0}^2} 
= v_\phi (R) \sqrt{\cos^2 i + \sin^2 i/M^4}$.
The post-shock velocity at position B is 
\bea
v_B \equiv v_{\xi 0}''(R+2d) = v_{\xi 0}(R+\Delta R) + \Delta v_\xi,  \label{eq: 6}
\eea
where $\Delta \xi =  {\Delta R}/{(R\sin i)}$, and
\bea
\Delta v_\xi &\equiv& \frac{\partial v_\xi (R+\Delta R)}{\partial \xi} \Delta \xi \\
& = & - (R + \Delta R) \frac{v_{\eta 1}}{v_{\eta 0}} \left. \frac{\kappa^2}{2 \Omega_c} \right|_{R+\Delta R} \frac{\tan i}{M^2} \frac{\Delta R}{R\sin i} \\
& \approx & 
- \frac{v_{\eta 1}}{v_{\eta 0}} \frac{\kappa^2}{2 \Omega_c} \frac{\Delta R}{M^2 \cos i} \;.
\label{eq: 13}
\eea
Here, we use 
$
{\partial v_{\xi }}/{\partial \xi} = {\partial v_{\xi }}/{\partial \eta} \cdot {\Delta \eta}/{\Delta \xi},
$
with
$ {\Delta \eta}/{\Delta \xi} =  \tan i/M^2 $,
and the linear solution of flow in a spiral potential (e.g. Spitzer 1978 [equation (13-24)]): 
\bea
\frac{\partial v_{\xi }}{\partial \eta} &\sim& \frac{\partial v_{\xi 1}}{\partial \eta} \\
 &=& - R \frac{v_{\eta 1}}{v_{\eta 0}} \frac{\kappa^2}{2 \Omega_c},
\eea
where $\Omega_c = v_\phi/R + \Omega_p$, $v_{\eta 0} \equiv v_\phi \sin i /M^2$,
and $v_{\eta 1}/v_{\eta 0}$ is given by
\bea
v_{\eta 1} \left[ -\left(\frac{2 v_\phi}{R}\right)^2 \left( 1 - \frac{c_s^2}{v_{\eta 0}^2}\right)
+ \kappa^2 \right] = -v_{\eta 0} \frac{\partial^2 \Phi_s}{R^2 \partial \eta^2},
\label{eq: 16}
\eea
with the epicyclic frequency $\kappa$ and a spiral potential $\Phi_s$.  Using equation (\ref{eq: 16}), equation (\ref{eq: 13})
is rewritten as 
\bea
\Delta v_\xi & \approx & 
\frac{ \frac{\partial^2 \Phi_s}{R^2 \partial \eta^2}}{
\left[ -(\frac{2 v_\phi(R)}{R})^2 \left( 1 - \frac{c_s^2}{v_{\eta 0}^2}\right)
+ \kappa^2 \right] 
} 
\frac{\kappa^2}{2 \Omega_c} \frac{\Delta R}{M^2 \cos i} \\
&=& 
\frac{ \frac{\partial^2 \Phi_s}{R^2 \partial \eta^2} \frac{\Delta R}{2 \Omega_c M^2 \cos i}}{
\left[ -2(\alpha + 1)^{-1} \left( 1 - M^2/\sin^2 i \right) + 1 \right] }
\; ,
\label{eq: 17}
\eea
where we assume a rotation curve as  $v_\phi \equiv v_0 (R/R_0)^\alpha$, 
therefore $\kappa^2 = 2 (\alpha + 1) (v_\phi/R)^2$, and
$v_{\eta 0} = v_\phi \sin i /M^2$ is used.
Using
$
\Phi_s(\eta) \equiv -\varepsilon \Phi_0 (R) \sin (2 \pi \eta / \pi \sin i),
$
 in the vicinity of a spiral shock,
\bea
\left.
 \frac{\partial^2 \Phi_s}{\partial \eta^2} \right|_{2d} = \frac{4\varepsilon  \Phi_0}{\sin^2 i} \sin\left(  \frac{2\eta}{\sin i} \right) \approx  \frac{4\varepsilon  \Phi_0}{\sin^2 i} \cdot \frac{2 \Delta \eta}{\sin i}, 
\label{eq: 18}
\eea
where $\Delta \eta =  {\Delta R}/{(R M^2\cos i)}$.
For a slow pattern speed, i.e. $v_\phi \sim R \Omega_c$, from
equations (\ref{eq: 17}) and (\ref{eq: 18})
then
\bea
\frac{\Delta v_\xi}{v_\phi} =  \mu \delta^2,
\label{eq: 19}
\eea
where $\delta \equiv \Delta R/R$, and 
\bea
\mu(M, i, \alpha, \varepsilon) \equiv \frac{
{4\varepsilon}/({M^4 \cos^2 i \cdot \sin^3 i})
}
{
\left[ 1 -2(\alpha + 1)^{-1} \left( 1 - M^2/\sin^2 i \right) \right]
}.
\label{eq: mu}
\eea
We finally have the velocity variation $u_0$ in equation (\ref{eq: richard_0}) 
using $\mu$ an $\delta$:
\bea
u_0^2 &= &  [(v_{\xi 0}''(R+2d) - v_{\xi 0}(R))/2 ]^2  \\
&=& \frac{1}{4}\left[ \frac{\partial v_\phi}{\partial R} \Delta R \cos i
 + \Delta v_\xi \right]^2  \\
&=&  \frac{v_\phi^2}{4}\left[ \alpha  \delta \cos i 
 + \mu\delta^2 \right]^2,
\label{eq: uzero}
\eea
where we use
$
v_{\xi 0} (R + \Delta R) - v_{\xi 0}(R) = [v_\phi(R+\Delta R) - v_\phi(R)] \cos i \, .
\label{eq: 27}
$

The density gradient behind the shock, $\sigma$, can be estimated using 
the mass conservation,  namely,
$
\rho_0 M^2 v'(R+\Delta R) = \rho'' v''(R+2d),
$
which is equivalent to 
\bea
\rho_0 M^2 v_\phi(R+\Delta R)(\sin^2 i/M^4 + \cos^2 i) = \nonumber \\ \rho'' v_{\xi 0}'' \left(1 + \frac{\tan^2 i}{M^4} \right)^{1/2},
\eea
where  $v_\eta/v_\xi = \tan i/M^2$ is used.
Therefore, the dimensionless density change in the layer, $\sigma$, is  
\bea
\sigma = \frac{\rho_0 M^2 - \rho''}{\rho_0 M^2 + \rho''} 
        & = & \frac{1 -\cos i \cdot {v_\phi}/{v_{\xi 0}''}} 
{1 + \cos i \cdot {v_\phi}/{v_{\xi 0}''}}  \\
&=& \frac{\alpha \delta \cos i+ \mu\delta^2}{2 \cos i + \mu\delta^2 + \alpha\delta \cos i} .
 \label{eq: epsilon}
\eea

The gravitational acceleration, transverse to the shock front, $g$, is given by
\bea
g &=& - \frac{\partial \Phi_s}{R\partial \eta}  \\
&=& \frac{v_\phi^2}{R} \left[
\frac{2 \varepsilon }{\sin i} \cos\left(\frac{2 \delta }{M^2 \cos i \sin i}\right)
\right]
\label{eq: gravi}
\eea



Finally, the Richardson number, $J$,
for the flow behind 
a spiral shock is given combining equations (\ref{eq: d}), (\ref{eq: uzero}), 
(\ref{eq: epsilon}), and (\ref{eq: gravi}) as
\bea
J(i, M, \varepsilon, \alpha; \hat{\lambda}) = \;\;\;\;\;\;\;\;\;\;\;\;\;\;\;\;\;\;\;\;\;\;\;\;\;\;\;\;\;\;\;\;\;\;\;\;\;\;\;\;\;\;\;\;\;\;\;\;\;\;\;\;\;\;\;\;\;\;\;\;\;\;\;\;\;\;\;\;\;\;\;\;\;\; \nonumber \\  \frac{
\frac{4 \varepsilon }{\sin i} \cos\left(\frac{2 \hat{\lambda}}{M^2 \cos i} \right)
} 
{M^2 \cos i (\alpha \cos i + \mu \hat{\lambda} \sin i)
({2 \cos i + \mu \hat{\lambda}^2 \sin^2 i+ \alpha \hat{\lambda} \cos i \sin i})
} \, .
\label{eq: final_richard}
\eea
where $\hat{\lambda}$ is a wavelength normalized by the radius,
$\hat{\lambda} \equiv \lambda/R  = \delta /\sin i$,
and $\mu$ is given by equation (\ref{eq: mu}).

\begin{figure}
\begin{center}
 	\includegraphics[width = 9cm]{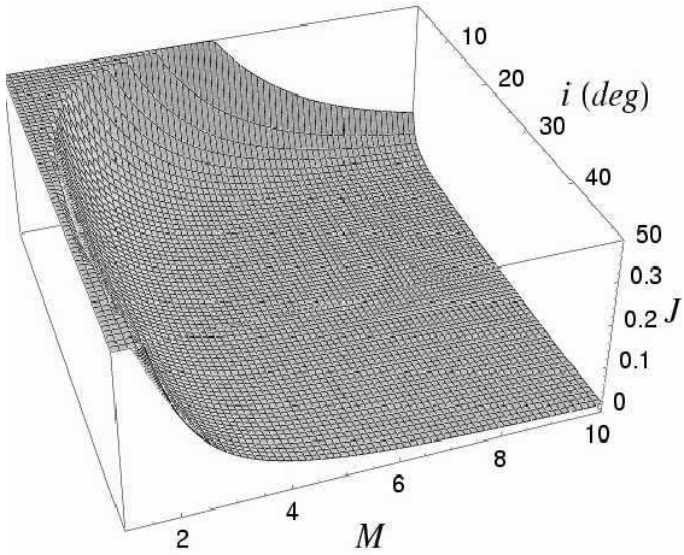}
        \caption{Richardson number,$J$, behind a spiral shock as a function of
Mach number, $M$, and the pitch angle of the spiral, $i$, for $\alpha = 1$ 
(i.e. rigid rotation ), $\hat{\lambda} = 0.1$,
and strength of the spiral potential, $\varepsilon = 0.1$.}
\label{fig: richard}
\end{center}
\end{figure}

%
%
In Fig. \ref{fig: richard},  the Richardson number given by 
equation (\ref{eq: final_richard})
is plotted as a function of the Mach number, $M$,  and the pitch angle, $i$.
It shows that $J$ monotonically decreases with $M$, therefore
weaker shock are expected to be more stable. For a given $M$, 
a larger $J$ is obtained for smaller $i$, if $i \la 20\degr$.
This suggests that tightly wound spiral shocks may be relatively stable.

Equation (\ref{eq: final_richard}) is approximated as
$
 J \approx 4\varepsilon/(i \alpha M^2),
$ for $i \ll 1$,
$\hat{\lambda} \ll 1 $ and $\alpha > 0$.
This suggests 
that spiral shocks in the region of a flat rotation (i.e. $\alpha = 0$) curve
are expected 
to be stable for short wavelengths in a galactic outer region.
In fact, one of the numerical results (Fig. \ref{fig: result_4})
shows that shocks in a flat rotation curve is stable. 
One should note, however, that 
even if the unperturbed circular velocity is
constant, a local gradient in circular velocity caused by
a spiral potential could destabilize the
spiral shocks on a local scale.

\begin{figure}
\begin{center}
 	\includegraphics[width = 7cm]{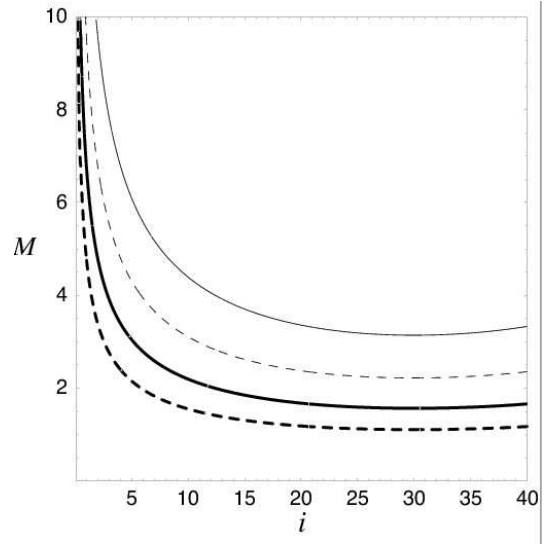}
        \caption{Curves for $J = 1/4$ are plotted as a function of the pitch angle $i$ (degree) and Mach number $M$, below which the spiral shocks could be
Kelvin-Helmholtz stable (i.e. $J > 1/4$, see Fig. \ref{fig: richard} and the text).
Thick curves are for rigid rotation ($\alpha = 1$), and thin curves are for Keplerian ($\alpha = -1/2$) with $\varepsilon = 0.1$ (solid curves) and 0.05 (dashed curves).
$\hat{\lambda} = $ 0.01 is assumed.
}
\label{fig: 4}
\end{center}
\end{figure}

In Fig. \ref{fig: 4}, the critical Richardson numbers 
are plotted as a function of the Mach numbers and the pitch angles for
two different rotation curves.
One should note again that $J > 1/4$ does not necessarily mean
that the sheared layer is K-H {\it stable}. This is only a
necessary condition for stability as mentioned in section 4.

The curves in Fig. \ref{fig: 4} have minima,
$M_{\rm min} \sim 3$ at pitch angles $i_{\rm min} \sim 30$. 
The plot shows that 
the $J =1/4$ curves decrease with the pitch angle, $i$, 
and they increase
for $i > i_{\rm min}$.
For $i < i_{\rm min}$, a flow is less affected by an
oblique shock with smaller pitch angles, therefore the velocity gradient
behind the shock is smaller, and as a result the shocked layer is expected to be more stable.
On the other hand, if the pitch angle is larger than $i_{\rm min}$,
the width of the sheared layer, $2d$, becomes larger,
and again the layer becomes stable.
We can also know that weak shocks with $M < M_{\rm min}$
are expected to be stable for all pitch angles.
$M_{\rm min}$ is larger in the Keplerian rotation curve than for the rigid
rotation. This means that spiral shocks generated around a point mass (e.g. an accretion disk
in a close-binary) tend
to be more stable than those generated in a potential with a constant mass density, for the same pitch angle and the Mach number.

For a given rotation curve, shocks may be more stable in 
stronger spiral potentials.
This is because the K-H instability can be
suppressed by buoyancy force, and stronger spiral potentials cause 
larger gravitational acceleration nearly perpendicular to the shock front 
[see equation (\ref{eq: richard_0})]. 
One should note however
that Mach number of the flow towards the potential minimum is
also affected by the strength of the potential, 
therefore it is not so straightforward to evaluate the effect of the spiral potential on
the wiggle instability.
If the spiral potential is too weak (e.g. amplitude of 
the spiral component is less than a few \% of the axisymmetric one), shocks do not appear in the flow (e.g. Shu et al. 1973).



In above discussion, we neglect the effect of expansion (perpendicular to 
the shock) velocity, $v_\xi$. In fact, the flow behind the shock
has the expansion velocity, which is approximately
\bea
\Delta v_\eta \approx (v_{\xi0}'' - v_{\xi 0}) \Delta \eta/\Delta \xi 
= 2 u_0 \tan i/M^2 = \frac{4 u_0}{M_0^2 \sin 2 i},
\label{eq: veta}
\eea
where $M_0 \equiv v_\phi/c_s$.
If $\Delta v_\eta^2 >  u_0^2$, we expect that the shear layer is stable for 
the K-H instability due to the expansion flow. From eq. (\ref{eq: veta}), 
it is suggested that the expansion velocity becomes important for spirals with
a small pitch angle and/or small Mach number (see also Dwarkadas \& Balbus 1996).
For example, this can be happen when $M_0 < $ 2.5, 3.4, and 4.8, for 
the spirals with $i = 20, 10,$ and 5$^\circ$, respectively.
In other words, for strong shocks with a large pitch angle, 
the post shock flow is nearly parallel to the shock, and then
the expansion velocity on the K-H instability can be negligible.
This is consistent with our numerical results.

\bsp
\label{lastpage}

\end{document}